\documentclass[12pt]{iopart}
\usepackage{graphicx}

\begin{document}

\title{Special functions, raising and lowering operators}
\author{N Cotfas}
\address{Faculty of Physics, University of Bucharest,
PO Box 76-54, Postal Office 76, Bucharest, Romania, 
E-mail address: ncotfas@yahoo.com}
\begin{abstract}
The Schr\" odinger equations which are exactly solvable in terms of
associated special functions are directly related to some self-adjoint
operators defined in the theory of hypergeometric type equations.
The fundamental formulae occurring in a supersymmetric approach to these
Hamiltonians are consequences of some formulae concerning the general
theory of associated special functions. We use this connection in order
to obtain a {\em general theory of Schr\" odinger equations exactly solvable
in terms of associated special functions}, and to extend certain results
known in the case of some particular potentials.
\end{abstract}

 \newcommand{\N}{\rm I\!N}
 \newcommand{\R}{\rm I\!R}

\section{Introduction}

It is well-known \cite{CKS,IH} that, in the case of certain potentials,
the Schr\" odinger equation is exactly solvable and its solutions
can be expressed in terms of the so-called {\it associated
special functions}. Our purpose is to present a general 
theory of these quantum systems. Our systematic study recovers
a number of earlier results in a natural unified way and also leads to
new findings.

The number of articles concerning exactly solvable quantum systems
and related subjects is very large (see \cite{CKS,IH,Jaf} and references
therein). Our approach is based on the 
raising/lowering operators presented in general form (for the first time 
to our knowledge) by Jafarizadeh and Fakhri \cite{Jaf}.
We reobtain these operators in a much simpler way, and use them in a 
rather different way. More details can be found in \cite{C}.

\section{Orthogonal polynomials and associated special functions}

Many problems in quantum mechanics and
mathematical physics lead to equations of hypergeometric type
\begin{equation}
\sigma (s)y''(s)+\tau (s)y'(s)+\lambda y(s)=0 \label{eq}
\end{equation}
where $\sigma (s)$ and $\tau (s)$ are polynomials of at most second
and first degree, respectively, and $\lambda $ is a constant. 
This equation can be reduced to the self-adjoint form 
\begin{equation}
[\sigma (s)\varrho (s)y'(s)]'+\lambda \varrho (s)y(s)=0 
\end{equation}
by choosing a function $\varrho $ such that 
$[\sigma (s)\varrho (s)]'=\tau (s)\varrho (s)$.
For $\lambda =\lambda _l=-\frac{1}{2}l(l-1)\sigma ''-l\tau '$
with $l\in {\N}$
there exists a polynomial $\Phi _l$ of degree $l$ 
satisfying (\ref{eq}), that is, 
\begin{equation}
\sigma (s)\Phi _l''(s)+\tau (s)
\Phi _l'(s)+\lambda_l \Phi _l(s)=0\, .  \label{eq1}
\end{equation}

If there exists a finite or infinite interval $(a,b)$ such that 
\begin{equation}\label{bounds}
\sigma (s)\varrho (s)s^k|_{s=a}=0\qquad 
 \sigma (s)\varrho (s)s^k|_{s=b}=0\qquad {\rm for\ all}\ \ k\in {\N}
\end{equation}
and if $\sigma (s)>0$, $\varrho (s)>0$ for all $s\in (a,b)$, then the
polynomials $\Phi _l$ are orthogonal with weight function $\varrho (s)$
in the interval $(a,b)$.
In this case $\Phi _l$ are known as 
{\em classical orthogonal polynomials} \cite{Nik}.

Let $\kappa (s)=\sqrt{\sigma (s)}$. 
By differentiating the equation (\ref{eq1}) $m$
times and multiplying it by $\kappa ^m(s)$, we get for
each $m\in \{ 0,1,2,...,l\} $ the 
{\em associated differential equation} which can be written as
$H_m\Phi_{l,m}=\lambda _l\Phi _{l,m}$, where
\[
\fl H_m=-\sigma (s) \frac{d^2}{ds^2}-\tau (s) \frac{d}{ds}
+\frac{m(m-2)}{4}\frac{{\sigma '}^2(s)}{\sigma (s)} 
+\frac{m\tau (s)}{2}\frac{\sigma '(s)}{\sigma (s)}
-\frac{1}{2}m(m-2)\sigma ''(s)-m\tau '(s) 
\]
and $\Phi _{l,m}(s)=\kappa ^m(s)\Phi _l^{(m)}(s)$
are known as the {\em associated special functions}. 
The set $\{ \Phi _{m,m},\Phi _{m+1,m},\Phi _{m+2,m},...\}$
is an orthogonal sequence (\cite{Nik}, pag. 8) in the Hilbert space
\[
\fl {\cal H}=\left\{ \varphi :(a,b)\longrightarrow {\R}\ \left|
\ \int_a^b|\varphi (s)|^2\varrho (s)ds<\infty \right. \right\} 
\quad {\rm with}\quad
\langle \varphi , \psi \rangle =\int_a^b\varphi (s)
{\psi }(s)\varrho (s)ds \, .
\]
For each $m\in {\N}$, let ${\cal H}_m$ be the linear span of 
$\{ \Phi _{m,m},\Phi _{m+1,m},\Phi _{m+2,m},...\}$. 
In the sequel we shall restrict us to the case when ${\cal H}_m$
is dense in ${\cal H}$ for all $m\in {\N}$. For this it is sufficient
the interval $(a,b)$ to be finite, but not necessary.

\section{Raising and lowering operators. Factorizations for $H_m$}

Lorente has shown recently \cite{L} that a factorization of $H_0$
can be obtained by using the well-known three term recurrence relation 
satisfied by $\Phi _l$ and a consequence of Rodrigues formula.
Following Lorente's idea we obtain a factorization of $H_m$ by
using the definition $\Phi _{l,m}(s)=\kappa ^m(s)\Phi _l^{(m)}(s)$ 
and a three term recurrence relation.

Differentiating $\Phi _{l,m}(s)=\kappa ^m(s)\Phi _l^{(m)}(s)$
we get the relation
\begin{equation}
\fl \Phi _{l,m+1}(s)=\left(\kappa (s)\frac{d}{ds}-
m\kappa '(s)\right) \Phi _{l,m}(s) \qquad {\rm for\ all\ }
m\in \{ 0,1,...,l-1\}.
\end{equation}
If we differentiate (\ref{eq1}) $m-1$ times and
multiply the obtained relation by $\kappa ^{m-1}$ then we get
for each $m\in \{ 1,2,...,l-1\}$ the three term recurrence relation
\begin{eqnarray}
\fl \Phi _{l,m+1}(s) + \left( \frac{\tau (s)}{\kappa (s)}
+2(m-1)\kappa '(s)\right)\Phi _{l,m}(s) 
+(\lambda _l-\lambda _{m-1}) \Phi _{l,m-1}(s)=0 
\end{eqnarray}
and $\left( \frac{\tau (s)}{\kappa (s)}+
2(l-1)\kappa '(s)\right) \Phi _{l,l}(s)
+(\lambda _l-\lambda _{l-1})\Phi _{l,l-1}(s)=0.$
A direct consequence of these formulae is the relation
\begin{equation}
\fl (\lambda _l-\lambda _m)\Phi _{l,m}(s)=
\left(-\kappa (s)\frac{d}{ds}-
   \frac{\tau (s)}{\kappa (s)}-(m-1)\kappa '(s)\right)
\Phi _{l,m+1}(s) 
\end{equation}
satisfied for all $m\in \{0,1,...,l-1\}.$

The operators 
$A_m:{\cal H}_m\longrightarrow {\cal H}_{m+1}$ and
$A_m^+:{\cal H}_{m+1}\longrightarrow {\cal H}_m$ defined by
\begin{equation}
\fl A_m=\kappa (s)\frac{d}{ds}-m\kappa '(s)\qquad 
A_m^+=-\kappa (s)\frac{d}{ds}-\frac{\tau (s)}{\kappa (s)}-(m-1)\kappa '(s)\, 
\end{equation}
satisfy the relations $A_m\Phi _{l,m}=\Phi _{l,m+1}$ and 
$A_m^+\Phi _{l,m+1}=(\lambda _l-\lambda _m)\Phi _{l,m}$ (see figure 1).
\begin{center}
\begin{figure}
\setlength{\unitlength}{1mm}
\begin{picture}(100,55)(-25,0)
\put(18.7,7){$a_0 $}
\put(11.8,7){$a_0^+$}
\put(28,18.3){$A_0^+$}
\put(28,12){$A_0$}
\put(25.2,8){$U_0$}
\put(18.7,22){$a_0 $}
\put(11.8,22){$a_0^+$}
\put(28,33.3){$A_0^+$}
\put(28,27){$A_0$}
\put(25.2,23){$U_0$}
\put(18.7,37){$a_0 $}
\put(11.8,37){$a_0^+$}
\put(28,48.3){$A_0^+$}
\put(28,42){$A_0$}
\put(25.2,38){$U_0$}
\put(43.7,22){$a_1 $}
\put(36.8,22){$a_1^+$}
\put(53,33.3){$A_1^+$}
\put(53,27){$A_1$}
\put(50.2,23){$U_1$}
\put(43.7,37){$a_1 $}
\put(36.8,37){$a_1^+$}
\put(53,48.3){$A_1^+$}
\put(53,42){$A_1$}
\put(50.2,38){$U_1$}
\put(68.7,37){$a_2 $}
\put(61.8,37){$a_2^+$}
\put(78,48.3){$A_2^+$}
\put(78,42){$A_2$}
\put(75.2,38){$U_2$}
\put(1,52){$.$}
\put(1,51){$.$}
\put(1,50){$.$}
\put(16,52){$.$}
\put(16,51){$.$}
\put(16,50){$.$}
\put(41,52){$.$}
\put(41,51){$.$}
\put(41,50){$.$}
\put(66,52){$.$}
\put(66,51){$.$}
\put(66,50){$.$}
\put(91,52){$.$}
\put(91,51){$.$}
\put(91,50){$.$}
\put(0,45){$\lambda _3$}
\put(0,30){$\lambda _2$}
\put(0,15){$\lambda _1$}
\put(0,0){$\lambda _0$}
\put(15,0){$\Phi _{0,0}$}
\put(15,15){$\Phi _{1,0}$}
\put(15,30){$\Phi _{2,0}$}
\put(15,45){$\Phi _{3,0}$}
\put(40,15){$\Phi _{1,1}$}
\put(40,30){$\Phi _{2,1}$}
\put(40,45){$\Phi _{3,1}$}
\put(65,30){$\Phi _{2,2}$}
\put(65,45){$\Phi _{3,2}$}
\put(90,45){$\Phi _{3,3}$}
\put(16,3){\vector(0,1){10}}
\put(18,13){\vector(0,-1){10}}
\put(22,15){\vector(1,0){17}}
\put(39,17){\vector(-1,0){17}}
\put(22,3){\vector(3,2){16}}
\put(16,18){\vector(0,1){10}}
\put(18,28){\vector(0,-1){10}}
\put(22,30){\vector(1,0){17}}
\put(39,32){\vector(-1,0){17}}
\put(22,18){\vector(3,2){16}}
\put(16,33){\vector(0,1){10}}
\put(18,43){\vector(0,-1){10}}
\put(22,45){\vector(1,0){17}}
\put(39,47){\vector(-1,0){17}}
\put(22,33){\vector(3,2){16}}
\put(41,18){\vector(0,1){10}}
\put(43,28){\vector(0,-1){10}}
\put(47,30){\vector(1,0){17}}
\put(64,32){\vector(-1,0){17}}
\put(47,18){\vector(3,2){16}}
\put(41,33){\vector(0,1){10}}
\put(43,43){\vector(0,-1){10}}
\put(47,45){\vector(1,0){17}}
\put(64,47){\vector(-1,0){17}}
\put(47,33){\vector(3,2){16}}
\put(66,33){\vector(0,1){10}}
\put(68,43){\vector(0,-1){10}}
\put(72,45){\vector(1,0){17}}
\put(89,47){\vector(-1,0){17}}
\put(72,33){\vector(3,2){16}}
\end{picture}
\caption{ The functions $\Phi _{l,m}$ satisfy the relation
$H_m\Phi _{l,m}=\lambda _l\Phi _{l,m}$, and are related
(up to some multiplicative constants) through the operators 
$A_m$, $A_m^+$, $a_m$, $a_m^+$, $U_m$ and $U_m^{-1}=U_m^+$. }
\end{figure}
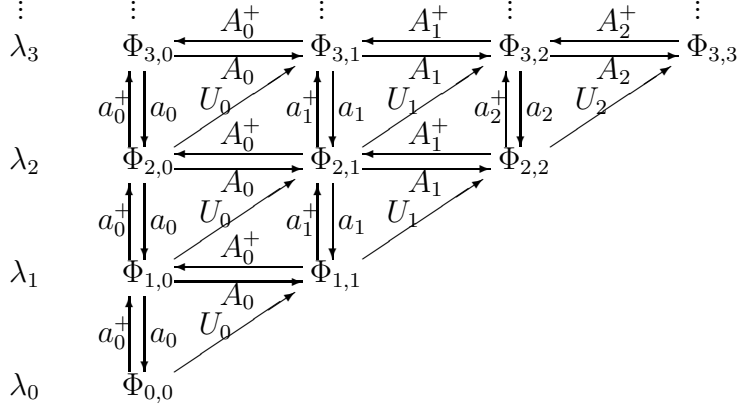
\end{center}
Since $\sigma ^m(s)\Phi _l^{(m)}(s)\Phi _k^{(m+1)}(s)$ is a polynomial and
the function $\sigma (s)\varrho (s)$ satisfies (\ref{bounds}),
integrating by parts one obtains $\langle A_m\Phi _{l,m},\Phi _{k,m+1}\rangle=
\langle \Phi _{l,m},A_m^+\Phi _{k,m+1}\rangle $, that is, the operators $A_m$
and $A_m^+$ are mutually adjoint. From the relation
\[
\fl ||\Phi _{l,m+1}||^2
=\langle \Phi _{l,m+1},\Phi _{l,m+1}\rangle
=\langle A_m\Phi _{l,m},\Phi _{l,m+1}\rangle 
=\langle \Phi _{l,m},A_m^+\Phi _{l,m+1}\rangle
=(\lambda _l-\lambda _m)||\Phi _{l,m}||^2 
\]
it follows that $\lambda _l>\lambda _m$ for all $l>m$, and
$||\Phi _{l,m+1}||=\sqrt{\lambda _l-\lambda _m}\, ||\Phi _{l,m}||.$
This is possible only if $\sigma ''(s)\leq 0$ and $\tau '(s)<0$.
Particularly, we have $\lambda _l\not= \lambda _k$ if and only if 
$l\not= k$.

The operators $H_m:{\cal H}_m\longrightarrow {\cal H}_m$ are self-adjoint, 
admit the factorizations
\begin{equation}
H_m-\lambda _m=A_m^+A_m\qquad H_{m+1}-\lambda _m=A_mA_m^+ 
\end{equation}
and satisfy the intertwining relations
$H_mA_m^+=A_m^+H_{m+1}$, $A_mH_m=H_{m+1}A_m.$

\section{Creation and annihilation operators. Coherent states}

For each $m\in {\N}$, the sequence
$\{ |m,m>,\, |m+1,m>,\, |m+2,m>,...\}$, where
$|l,m>=\Phi _{l,m}/||\Phi _{l,m}||$
is an orthonormal basis of ${\cal H}$, and
$U_m:{\cal H}\longrightarrow {\cal H}$, 
$U_m |l,m\rangle =|l+1,m+1\rangle $
is a unitary operator.
The mutually adjoint operators (see figure 1)
$a_m,\, a_m^+:{\cal H}_m\longrightarrow {\cal H}_m$,   
$a_m=U_m^+A_m$, $a_m^+=A_m^+U_m$
satisfy the relations
\begin{equation}
\fl a_m|l,m\rangle =
\sqrt{\lambda _{l}-\lambda _m}\, |l-1,m\rangle \qquad 
a_m^+|l,m\rangle = 
\sqrt{\lambda _{l+1}-\lambda _m}\, |l+1,m\rangle 
\end{equation}
and allow us to factorize $H_m$ as $H_m-\lambda _m=a_m^+a_m .$

The Lie algebra ${\cal L}_m$ generated by 
$\{ a_m^+,a_m \}$ is isomorphic to $su(1,1)$ if $\sigma ''<0$, and it is
isomorphic to the Heisenberg-Weyl algebra $h(2)$ if $\sigma ''=0.$

Let $m\in {\N}$ be a fixed natural number, and let
$|n\rangle =|m+n,m\rangle$, $e_n=\lambda _{m+n}-\lambda _m$,
$\varepsilon _0=1$, $\varepsilon _n=e_1e_2...e_n$. 
Since $0=e_0<e_1<e_2<...<e_n<...$
and
\begin{equation}
\fl a_m|n\rangle =\sqrt{e_n}\, |n-1\rangle \qquad 
a_m^+|n\rangle =\sqrt{e_{n+1}}\, |n+1\rangle \qquad 
(H_m-\lambda _m)|n\rangle =e_n|n\rangle
\end{equation}
we can define a system of coherent states by using the general setting
presented in \cite{AG}.

If $R=\limsup_{n\rightarrow \infty } \sqrt[n]{\varepsilon _n} \not= 0$
then we can define
\begin{equation}
|z\rangle = \frac{1}{N(|z|^2)}\sum_{n\geq 0}
\frac{z^n}{\sqrt{\varepsilon _n}}|n\rangle \qquad {\rm where} \qquad
(N(|z|^2)^2=\sum_{n=0}^\infty \frac{|z|^{2n}}{\varepsilon _n}
\end{equation}
for any $z$ in the open disk $C(0,R)$ of center $0$ and radius $R$.
We get in this way a continuous family 
$\{ |z\rangle |\, z\in C(0,R)\, \}$ of normalized coherent states
such that $a_m|z\rangle =z|z\rangle .$

\section{Application to Schr\" odinger type operators}

The problem of factorization of operators $H_m$ is a very important one
since it is directly related to the factorization of some Schr\" odinger 
type operators \cite{CKS,IH}.
If we use a change of variable $s=s(x)$ 
such that $ds/dx=\kappa (s(x))$ or $ds/dx=-\kappa (s(x))$ and
define the new functions 
$\Psi _{l,m}(x)=\sqrt{\kappa (s(x))\, \varrho (s(x))}\, \Phi _{l,m}(s(x))$
then  the relation $H_m\Phi_{l,m}=\lambda _l\Phi _{l,m}$  
becomes an equation of Schr\" odinger type
\begin{equation}
-\frac{d^2}{dx^2}\Psi _{l,m}(x)+V_m(x)\Psi _{l,m}(x)
=\lambda _l\Psi _{l,m}(x) .
\end{equation}

For example, by starting from the equation of Jacobi polynomials with 
$\alpha =\mu -1/2$, $\beta =\eta-1/2$, and using the change of
variable $s(x)=\cos x$ we obtain the Schr\" odinger equation 
corresponding to the P\" oschl-Teller potential \cite{AG}
\begin{equation}\label{PT}
 V_0(x)=\frac{1}{4}\left[ \frac{\mu (\mu -1)}{\cos ^2(x/2)}+
\frac{\eta (\eta -1)}{\sin ^2(x/2)}\right]-\frac{(\mu +\eta )^2}{4}.
\end{equation}

If we choose the change of variable $s=s(x)$ such that 
$ds/dx=\kappa (s(x))$, then the operators corresponding to 
$A_m$ and $A_m^+ $ are the adjoint conjugate operators
\begin{equation}
\begin{array}{l}
{\cal A}_m=[\kappa (s)\varrho (s)]^{1/2}A_m
[\kappa (s)\varrho (s)]^{-1/2}|_{s=s(x)}
=\frac{d}{dx}+W_m(x)\\[2mm]
{\cal A}_m^+ =[\kappa (s)\varrho (s)]^{1/2}A_m^+ 
[\kappa (s)\varrho (s)]^{-1/2}|_{s=s(x)}
=-\frac{d}{dx}+W_m(x)
\end{array}
\end{equation}
where the {\em superpotential} $W_m(x)$ is given by the formula
\begin{equation}
W_m(x)=-\frac{\tau (s(x))}{2\kappa (s(x))}
-\frac{2m-1}{2\kappa (s(x))}\frac{d}{dx}\kappa (s(x))\, .
\end{equation}

From the relations satisfied by $A_m$ and $A_m^+$ we get 
the formulae
\begin{equation}
\fl -\frac{d^2}{dx^2}+V_m(x)-\lambda _m={\cal A}_m^+ {\cal A}_m\qquad
-\frac{d^2}{dx^2}+V_{m+1}(x)-\lambda _m={\cal A}_m{\cal A}_m^+ 
\end{equation}
\begin{equation}
\fl V_m(x)-\lambda _m=W_m^2(x)-\dot W_m(x)\qquad
V_{m+1}(x)-\lambda _m=W_m^2(x)+\dot W_m(x)
\end{equation}
\begin{equation}
W_m(x)=-\frac{\dot \Psi _{m,m}(x)}{\Psi _{m,m}(x)}\qquad \qquad
V_m(x)=\frac{\ddot \Psi _{m,m}(x)}{\Psi _{m,m}(x)}+\lambda _m 
\end{equation}
where the dot sign means derivative with respect to $x$. 

If we choose the change of variable $s=s(x)$ such that 
$ds/dx=-\kappa (s(x))$ then the formulae are very similar 
(only some signs are changed). In the case of P\" oschl-Teller 
potential (\ref{PT}) we get
\[ W_0(x)=\frac{1}{2}\left[ \mu \cot \frac{x}{2}
-\eta \tan \frac{x}{2}\right]. \]

\end{document}